\documentstyle[12pt]{article}

\newcommand{\rf}[1]{(\ref{#1})}

\renewcommand{\thefootnote}{\fnsymbol{footnote}}

\newcommand{\newsection}{    % Numeration of eqs. is automatic
\setcounter{equation}{0}
\section}
\def\appendix#1{
  \addtocounter{section}{1}
  \setcounter{equation}{0}
  \renewcommand{\thesection}{\Alph{section}}
  \section*{Appendix \thesection\protect\indent \parbox[t]{11.715cm} {#1} }
  \addcontentsline{toc}{section}{Appendix \thesection\ \ \ #1}
  }

\newcommand{\eq}[1]{Eq.~(\ref{#1})}
\newcommand{\tr}[1]{\:{\rm Tr}\,#1}

\def \up {\uparrow}

\def\bi {\bibitem}

\def \om {\omega}

\def \F {{\cal F}}
\def \g {\gamma}
\def \del {\partial}

\def \a {\alpha}

\def \chi {\chi}

\def \m {\mu}
\def \n {\nu}

\def \t {\tau}
\def \td {\tilde }
\def \d {\delta}
\def \ci {\cite}

\def \P {\Phi}

\def \inv {^{-1}}
\def \ov {\over }

\def \fourth{{{1\over 4}}}

\def \e { e }
\def \t {\tau}
\def \V {{\cal V}}
\def \up {\uparrow}
\def \pa {\Vert}
\def \hal{{{1\over 2}}}

 \def \G {{\cal G}}

 \def \P {{\cal P}}
 \def \d {\delta}

\def \G {{\cal G}}
\def \foot{\footnote}

\def\np {{  Nucl. Phys. }}
\def \pl {{  Phys. Lett. }}
\def \mpl {{ Mod. Phys. Lett. }}

\def \cmp {{ Commun. Math. Phys. }}
\def \ijmp {{ Int. J. Mod. Phys. }}

\def \bi{\bibitem}

\def \T {{\cal T}}
\def\det{\hbox{det}}
\def\be{\begin{equation}}
\def\ee{\end{equation}}
\def\beq{\begin{equation}}
\def\eeq{\end{equation}}
\def\bea{\begin{eqnarray}}
\def\eea{\end{eqnarray}}
\def\LB{\left (}
\def\RB{\right )}
\def\lsb{\left [}
\def\rsb{\right ]}
\def\te{\theta}
\def\de{\partial}
\def\deu{\partial_{\tau}^2}
\def\xo{{\bar X}^{(1)}}
\def\xt{{\bar X}^{(2)}}
\def\xijo{{\bar X}^{(1)}_{ij}}
\def\xijt{{\bar X}^{(2)}_{ij}}
\def \la{\label}
\def \ci{\cite}
\def \fo {$4\pa 0\ $}
\newcommand{\non}{\nonumber \\*}
\begin{document}
\begin{titlepage}
\begin{flushright}
ITEP--TH--12/97\\
ITP-SB-97-23\\
Imperial/TP/96-97/42\\
hep-th/9704127\\
\end{flushright}
\vspace{.5cm}

\begin{center}
{\LARGE  Long-distance  interactions of D-brane  bound states  \\[.3cm]
and longitudinal 5-brane in M(atrix) theory   }\\[.2cm]
\vspace{1.1cm}
{\large I. Chepelev${}^{{\rm 1,}}$\footnote{E-mail: guest@vxitep.itep.ru}
and A.A. Tseytlin${}^{{\rm 2,}}$\footnote{Also at Lebedev Physics
Institute, Moscow. \ E-mail: tseytlin@ic.ac.uk} }\\
\vspace{18pt}
${}^{{\rm 1\ }}${\it Institute of Theoretical and Experimental Physics,
Moscow 117259, Russia}\\
${}^{{\rm 2\ }}${\it ITP, SUNY at Stony Brook, NY
 and  Imperial College, London SW7 2BZ, U.K.}
\end{center}
\vskip 0.6 cm

\begin{abstract}
We discuss long-distance, low-velocity interaction potentials for
processes involving longitudinal M5-brane (corresponding in type
IIA theory language to the 1/4 supersymmetric bound state of
4-brane and 0-brane). We consider the following scattering configurations:
(a) $D=11$ graviton off longitudinal M5-brane,
or, equivalently, 0-branes off marginal $4\pa 0$ bound state;
(b) M2-brane off longitudinal M5-brane, or, equivalently,
a non-marginal $2+0$ bound state off  marginal $4\pa 0$ bound state;
(c) two parallel longitudinal M5-branes, or, equivalently,
two $4\pa 0$ marginal bound states. We demonstrate the
equivalence between the classical string theory (supergravity)
and M(atrix) model (one-loop super Yang-Mills) results
for the leading terms in the interaction potentials. The supergravity
results are obtained using a generalisation of a classical
probe method which allows one to treat bound states of
D-branes as probes by introducing non-zero world-volume gauge field
backgrounds.
\end{abstract}

\end{titlepage}
\setcounter{page}{1}
\renewcommand{\thefootnote}{\arabic{footnote}}
\setcounter{footnote}{0}

%%%%%%%%%%%%%%%%%%%%%%%%%%%%%%%%%%%%%%%%%%%%%%%%%%%%
%%%%%%%%%%%%%%%%%%%%%%%%%%%%%%%%%%%%%%%%%%%%%%%%%%%%%%%%%

\newsection{Introduction}
%%%%%%%%%%%%%%%%%%%%%%%%%%%%%%%%%%%%%%%%%%%%%
The  M(atrix) theory proposal \ci{bfss} that the   dynamics
of the $D=11$  M-theory \cite{wit1}
at large values of 11-dimensional momenta
can be described by large $N$  10-dimensional $U(N)$ Super Yang-Mills
theory  reduced to
 1+0
dimensions  (which
represents the
dynamics  of a large number of  0-branes \cite{pol,wit2} at
short distances and low velocities)
was studied in a number of recent
papers. When $p$ of spatial  directions are compactified on a torus
 the $U(N)$
 SYM
quantum mechanics is to be replaced by   $U(N)$  SYM  field theory
reduced to $1+p$ dimensions  \cite{bfss,tay,sus,grt,dvv}.

Various p-brane configurations  appear in this approach as classical
operator solutions of  the $U(N)$ SYM theory \ci{bfss,grt,bss}.
The long-distance interactions of some  of such  1/2 supersymmetric
 p-branes were studied
\ci{ab,lifmat,lif3,corr} and were shown to be in agreement with supergravity
predictions.\foot{A  non-trivial generalisation to case of
non-zero $p_{11}$ transfer was considered in \ci{pp}.}
The  presence of the large number of 0-branes or, equivalently,
of the large boost in 11-th dimension,  implies that
these  branes  are not `pure' but, from type IIA theory point of view,
represent 1/2 supersymmetric  non-marginal bound states of p-branes with
0-branes (and, in general, other  branes) \ci{lifmat,lif3}.
For example, the 2-brane of M(atrix) theory \ci{nic,bfss}
 corresponds to
a type IIA 2-brane `populated' by  a  large  number
0-branes, i.e.   to
 the 2+0 non-marginal bound state \ci{pol}.
  This interpretation is  consistent with the fact that
 the classical  $D=11$ 2-brane solution
 boosted in 11-th direction has the
 2+0 configuration as its
 dimensional reduction \ci{rut}.

  The 2+0 bound state
 can be described as a 2-brane with a
 constant  magnetic  field   on its world volume (the 2-brane
 couples to  the RR vector  field
 via CS term  $\int C_1 \wedge { F} $  \cite{doug} and thus
  the induced 0-brane
 charge is  $N= {1\ov 2\pi} \int F$). The  value of the  magnetic field
 is thus  directly related to the   value of the boost in 11-th dimension
 \ci{lifmat}. In general,  the presence of a magnetic flux on a
 Dp-brane
 producing a 0-brane charge  leads also
 to the presence of other RR charges  on the brane
 describing a non-marginal bound state which is $O(d,d)$ T-dual
 to the pure Dp-brane. This     was discussed  for   
  `4+2+0'-brane\foot{From M-theory point of view, this 
  is  a 1/2 supersymmetric  non-marginal 
 bound state of longitudinally wrapped and boosted  5-brane 
 and two orthogonal  2-branes \ci{bss}.
  For  corresponding classical solution see \ci{ts4}.} and 
  `6+4+2+0'-brane in   \ci{lif2,lif3}.
  For large value of $N$ or  large  magnetic  flux
  certain configurations of  supersymmetric  (bound states of) p-branes
  become nearly BPS   and thus  several  ($v^n$, $n\leq 4$)  leading terms
  in their low-velocity interaction potentials  computed using
  open
  string theory description \ci{pol}
  have the same short-distance and large-distance
  forms  \ci{dkps,lif2}.
  Since the M(atrix) theory  description  is essentially equivalent
  to the open-string description at short distances
  (where only  the light open string
  modes are the  relevant degrees of freedom),
  this explains \ci{bfss,lifmat}
  why the  M(atrix)  theory  approach   should
   match the long-distance   supergravity
    description.

The aim of the present paper is to extend the checks done in
\ci{ab,lifmat,lif3}
%,ballar,corr}
  to  interactions involving
  another type of  M-brane  configuration  --
the 1/4 supersymmetric marginal   bound state  of    M5-brane
  and    momentum  wave  in 11-th direction, or
$5_M \pa \up$.\foot{We  shall use the notation $p\pa q$
for a marginal bound state of a $p$-brane and a $q$-brane
and $p+q$ for a non-marginal one (for a review, see \ci{ts4}).}
This configuration may be  viewed either as
 the  extremal limit of the black
 M5-brane \ci{guv} infinitely boosted in  the
longitudinal (11-th) direction, or as a
 BPS state of M5-brane
with additional transverse left-moving oscillations
carrying 11-dimensional momentum.
The corresponding  1/4 supersymmetric $D=11$ supergravity solution
 is parametrised  by two independent charges, or, more generally,
 two harmonic functions \ci{ts2}.
Upon dimensional reduction  along  11-th dimension
it becomes the 1/4 supersymmetric  $4\pa 0$ type IIA
configuration,  representing   the  marginal bound state
of the 4-brane and 0-brane
(the corresponding classical type IIA solution
\ci{ts2}
is U-dual to $5 \pa 1$ background   \ci{ts1}).

The matrix model  description   of the  $5_M \pa\up$
configuration  in the  case
when momentum flow is along 11-th direction
 (`longitudinal 5-brane')
  was suggested  in
 \ci{grt,bss}.\foot{For  discussions
 of   transverse M5-brane see \ci{grt,lif4,others}.}
One  considers the  self-dual configurations on $\td T^4$
(we shall assume that 5-brane is  wrapped over
 a 4-torus)
$ F_{ab} =*F_{ab}   , \  \
  F_{ab} = -i T^{-2} [X_a,X_b] , \  \
   X_a= T\inv (i\del_a + A_a )$  , $\ T= (2\pi
  \a')\inv$,
  where, in the case of a   single  5-brane  $\int_{\td T^4} d^4x {\
  \rm tr} (F_{ab}*F_{ab}) = 16 \pi^2 $.
  This is in direct correspondence with the description of
$4\pa 0$  as a 4-brane with a 4-d gauge instanton background in
its 1+4 world volume \ci{wit3,doug}.
As was  explained in \ci{tay,grt}, the  presence of the
winding strings in the case of  a D-brane on a torus  is  represented
in the  T-dual  picture of  large $N$
 SYM theory on  dual torus  by  the configuration  with $X_a$
equal to the 
  covariant derivative
operators,
$ T\inv ( i\del_a  + A_a)$.

Our aim will be to  test this description
by demonstrating the equivalence of
the  long-distance, low-velocity   interaction
potentials for longitudinal 5-brane  in the one-loop
 SYM description
and in the  corresponding  classical closed
string theory (supergravity) description.
The  new element  compared to the
 previous similar tests in    \ci{ab,lifmat,lif3}
is that
here we deal with  interactions of 1/4
supersymmetric  objects, which, in  the M(atrix) theory
description,  are in general represented by  {non-abelian}
YM  backgrounds.

 %%%%%%%%%%%%%%%%%%%%%%%%%%%%%%%%%%%%%%%%%%%%%%%

 We  shall  demonstrate the equivalence
 between the  supergravity and  SYM results
 for   the following   configurations:

 (a)  scattering of a $D=11$ graviton on a longitudinal M5-brane,
 or,  in type IIA language,
    scattering  of a
 bound state of 0-branes  on  marginal $4\pa 0$   bound state;

 (b)  scattering of  M2-brane  off longitudinal M5-brane,
 or, equivalently,  scattering of  a non-marginal $2+0$
 bound state  on  a marginal  $4\pa 0$ bound state;

 (c) scattering of two parallel  longitudinal M5-branes,
 or, equivalently,   scattering of  two $4\pa 0$ marginal
  bound states.

 In the first and the third  cases  the force 
 vanishes  for zero velocity (the static configurations are BPS)
 and we will be interested in the leading  velocity-dependent
 $v^2$ and $v^4$ terms in the  long-distance interaction
 potential.
 In the second case we have chosen to
  consider the orthogonal orientation
 of the  2-brane and  the 5-brane which is not BPS
  (the BPS configuration
 corresponds to the case of one common dimension \ci{ts3})
 and thus the leading term in the potential
 will be velocity-independent.
  This configuration    becomes  approximately BPS
 in the limit of  infinite boost or
 large number of 0-branes,
 allowing one to  establish  the equivalence  between the
 supergravity and M(atrix) theory results in a similar way as in
 \ci{lifmat}.
 The analysis of other possible configurations is similar.

  To determine  the   interaction
  potentials
 of these brane systems
 as described by  the massless sector of the classical
 closed  string theory (supergravity)
 we shall use  the classical probe method.
 As we shall explain below, it
  gives a simple  universal way of  deriving
 not only  the  low-velocity but  also
  the full relativistic
 expressions
  for the long-distance  brane scattering
   (in the special case of   0-brane scattering a different
   method was discussed
    in \ci{ballar}).

 The basic idea of the classical  probe method
  \ci{calkhu,duff,ts3,ts4,dps}
  is to consider the motion
 of a  probe -- a    p-brane (or a bound state of branes)
 in a background produced by  a source -- a q-brane
 (or a bound state of  branes).
  The information  which is used is
 essentially the one encoded in the supergravity action:
 (i) the form of the backgrounds produced by p-brane sources
 and their composite configurations
 and (ii) the form of the actions for  collective coordinates
 of such solitons.
 In principle, the knowledge of the
 supergravity action  (low-energy closed string theory effective
   action) and its p-brane solutions  is  enough
 to deduce the form of the  probe actions,
 but it is usually  much simpler
 to  invoke    string-theory  considerations
   to determine the detailed structure of the
 couplings to external fields
 (in particular, to use the relation to open string
 theory  \ci{pol,lei}  and T-duality
 to fix the form of the Born-Infeld term  \ci{FT}
  in D-brane  action \ci{bachas,ts5}
  and the  Chern-Simons couplings  to external RR  fields \ci{doug}).

%%%%%%%%%%%%%%%%%%%%%%%%%%%%%%%%%%%%%%%%%%%%%%%%%%%%%%%%%%

The new element in the present discussion  will be the
 treatment of   bound states of D-branes as classical  probes.
As we shall explain,  for this one needs
to  extend the probe method by  introducing 
  non-zero  world-volume gauge field
 backgrounds in the classical  probe actions.
 We shall
demonstrate that this gives a simple way of computing the
long-distance  classical relativistic
 scattering potentials  for   (marginal and non-marginal)
 BPS bound states of D-branes.
  %%%%%%%%%%%%%%%%%%%%%%%%%%%%%%%%%%%%%%%%%%%%%%

While our specific  results about the agreement of  the
first two leading-order
terms in the potentials of scattering  processes involving
longitudinal 5-brane
computed in the M(atrix) theory and in supergravity are not unexpected
(all configurations we discuss have approximate
supersymmetry for small velocities and/or large magnetic fluxes),
  our methods   have   wider  applicability.
In particular, the  probe method applies
to
various types of  composite BPS states of branes, and the
large-mass  expansions of the one-loop SYM
effective action discussed  below
 allow one to study
the scattering of these branes from M(atrix) theory point of view.

In the first part of the paper we shall  determine
the classical closed string (supergravity)
 results for the interaction potentials.
We shall start in section 2.1 with a
description  of  the probe method  with non-trivial world-volume
gauge field fluxes.
  We shall consider in detail  the example of the  0-brane scattering off
 $2+0$  bound state, treating the latter as a probe, and the former as
 a source.
   We shall reproduce  in a simple way
 the full relativistic
 expression for  the  scattering potential found previously in
 \ci{lif2,ballar}.
 In section 2.2 we shall apply  this method to the case of
 $(2+0)$  -- $(4\pa 0)$ bound state interaction. The case
 of 0-brane scattering off $4\pa 0$ bound state will be
analysed   in section 2.3.
 In section 2.4 we shall explain how
 to  describe the  $4\pa 0$ state   as a probe by using the
 4-brane  action with a constant self-dual
magnetic  gauge field background on a 4-torus.  We shall then compute the
interaction potential between the two $4\pa 0$ bound states
which will turn out to have the expected  `probe -- source'
symmetry.

In the  second part of the paper will present the corresponding
M(atrix) model calculations and demonstrate  their equivalence with
the supergravity results. The general structure of the one-loop SYM
effective action will be discussed
in section 3.1  and Appendix A.  In section 3.2
we shall determine the phase shift for the graviton -- longitudinal
5-brane scattering (the details  of computation will be given in
 Appendix B).  In section 3.3 we shall  find  the static potential
 between the orthogonally oriented membrane and longitudinal 5-brane.
  The computation of the scattering phase shift
  in the 
 case of  two  slowly moving
  parallel longitudinal 5-branes (section 3.4)  is similar   and 
  its result  also agrees
  with the supergravity result   found in section 2.4.

%%%%%%%%%%%%%%%%%%%%%%%%%%%%%%%%%%%%%%%%%%%%%%%%%%%%%%%%%%
\newsection{Long-distance interaction potentials: 
 closed string  effective  field  theory   (supergravity)  description }
%%%%%%%%%%%%%%%%%%%%%%%%%%%%%%%%%%%%%%%%%%%%%%%%%%%%%%%%%%
%%%%%%%%%%%%%%%%%%%%%%%%%%%%%%%%%%%%%%%%%%%%%%%%%%%%%%
\subsection{Classical probe method: $0$ - $(2+0) $ interaction}
%%%%%%%%%%%%%%%%%%%%%%%%%%%%%%%%%%%%%%%%%%%

To illustrate the probe method we shall be using to obtain
long-distance, low energy  classical   closed  string theory  
 interaction potentials
let us first consider the case of  a $D=11$ graviton scattering on (boosted)
M2-brane, or, equivalently, the  0-brane  scattering  off  $2+0$
bound state, which was  discussed
previously in \ci{lif2,ab,lifmat,ballar}.

We shall  consider the $2+0$ probe moving in the 0-brane background.
The same  result is  found
by  studying the 0-brane probe in  the 2+0 background
 \ci{rut}
  but we are interested here in clarifying how
the $2+0$ probe can be described  by  a 2-brane  action
 with a constant magnetic field background.

 Let us assume that the  2-brane   is  wrapped over
 a 2-torus $(R_1,R_2)$  with volume $V_2=(2\pi)^2 R_1 R_2$
  and that the 0-brane  background
 is  `smeared' in these directions.
 The  relevant terms in   the
  action  for a  2-brane  probe propagating in  curved  space
   are   ($m,n=0,1,2; \ i,j=3,...,9$)
 \be
 I_2 = - T_2\int d^3 x  \bigg[  e^{-\phi} \sqrt {-\det\big( G_{mn } +
 G_{ij} \del_m X^i \del_n X^j +  \F_{mn} \big)}
    - {1\over 6} e^{mnk} C_{mnk}  -  {1\over 2}  \e^{mnk} C_m \F_{nk}  \bigg] \ ,
    \la{act}
    \ee
    where $\F_{mn} \equiv  T\inv  F_{mn} + B_{mn} $
   and
    $C_m, C_{mnk}$ are  the RR fields
     (in this section $B_{mn}=0, \ C_{mnk}=0$).
      We used the static gauge ($X_m=x_m$)
    and assumed
    that the metric has 3+7 block-diagonal form.
    In general, Dp-brane tension is \ci{pol}
   \be
  T_p \equiv n_p \bar T_p=  n_p g\inv (2\pi)^{(1-p)/2}  T^{(p+1)/2} \ , \ \ \ \ \ \
   T\equiv (2\pi \a')^{-1} \ ,
  \la{tens}
  \ee
   so that  the tension of  2-brane  with charge $n_2$   is
  $\  T_2 = n_2 g\inv (2\pi)^{-1/2}  T^{3/2} .$

  The type IIA  0-brane background (`smeared' in  two  directions
  $x_1=y_1,\ x_2=y_2$)  is  \ci{HS,duff}
   \be
   ds^2_{10} = H_0^{1/2}  [ - H_0^{-1}dt^2 + dy_1^2 + dy_2^2 +
    dx_i dx_i ] \ ,
    \la{two} \ee
$$ e^{\phi} = H_0^{3/4} \ , \ \ \ \ \  C_0 = H_0^{-1}-1  \ , \ \ \ \
H_0 = 1 + {Q^{(2)}_0\over r^5} \ , \ \ \ \ r^2=x_i x_i\  .    $$
We shall use the notation $Q^{(n)}_p$  for the coefficient in the harmonic
function $H_p = 1 + {Q^{(n)}_p\over r^{7-p-n}}$
of p-brane  background which is smeared in $n$ transverse toroidal directions.
In general,
\be
Q_p =  N_p g (2\pi)^{(5-p)/2} T^{(p-7)/2} \om_{6-p}\inv  \ , \ \ \  \ \
 \om_{k-1} = 2 \pi^{k/2}/\Gamma({ k/ 2})  \ ,
\la{qq}
\ee
\be
Q^{(n)}_p = N_p g (2\pi)^{(5-p)/2} T^{(p-7)/2} (V_n \om_{6-p-n})\inv
   , \    \
{\rm i.e.}     \
\
Q^{(n)}_p = N_p N_{p+n}\inv  Q_{p+n} (2\pi)^{n/2} T^{n/2} V_n\inv  ,
\la{sme}
\ee
where  $V_n$ is the  volume of the flat internal  torus.
Thus
$
Q^{(2)}_0 =  N_0 g (2\pi)^{5/2}    T^{-7/2}
 (V_2  \omega_{4})\inv   ,\
 $
where  $N_0$ is the  charge of the 0-brane  source
($\omega_{4}= {8\ov 3} \pi^2, \  \omega_{6}= {16\ov 15} \pi^3 $).

To describe  the classical  probe  which represents
  the  motion of the supergravity soliton  corresponding to the
   $2+0$  bound state  with $n_2$ units of 2-brane charge and $n_0$ units of
   0-brane charge
we shall assume, following \ci{lif2},  that
the $U(1)$ gauge field $F_{mn}$ has a non-trivial  constant
magnetic  background $\F_{12} =T\inv F =  \F $ such that
\be
 T_2 \int_{T^2}  \F = T_2 V_2 \F =  T_0   \ , \ \ \ \ \
     \F  = 2\pi  (V_2 T)\inv { n_0 \ov n_2 }
      = (2\pi)\inv  \td V_2 T  { n_0 \ov n_2 } = { n_0 \ov n_2 } \ .
\la{inte}
\ee
 Here $\td V_2 =(2\pi)^2 (T^2V_2)\inv $
  is the volume of T-dual torus  and  we have assumed that
 $V_2$   has self-dual
 radii,  $R_i=\sqrt {\a'}$, so that
 $2\pi  (V_2 T)\inv  =(2\pi)\inv  \td V_2 T =1$.
  T-duality along the two toroidal directions then
   corresponds to interchanging  $n_0$ and $n_2$. 
   To simplify the expressions,  in what follows
    we shall  always consider tori with self-dual radii, for which
   \be
   V_n = (2\pi)^{n/2} T^{-n/2} \ , \ \ {\rm i.e.} \ \ \   T_p V_p=n_p
   n_0\inv  T_0\ , \ \ \
   Q^{(n)}_p = N_p N_{p+n}\inv  Q_{p+n} \ .
   \la{self}
   \ee
   This assumption is not necessary
   and  the volume factors can be easily restored
   using  the general form of $\F$ in  \rf{inte}.

  Substituting  the  background  \rf{two}  into the  2-brane action and assuming that
$X_i$ depend only on $t$, i.e.
 keeping   only the velocity ($v_i = \del_0 X_i$) dependent terms
 in the action,
 we find that  the matrix under the  square root in \rf{act}  is
$$
\pmatrix{  H_0^{-{1\over 2} } (1  - H_0 v^2)   &   0 & 0 \cr
            0     &  H_0^{1/2} & \F \cr
            0    &   -\F    &    H_0^{1/2}  \cr} \ ,  $$
 so that
      \be
 I_{2+0}  = -T_2 V_2 \int dt  \bigg[ H_0^{-1}   \sqrt { (1-  H_0 v^2)( \F^2 + H_0 ) }
    - (H_0^{-1} -1)\F  \bigg]  \ ,
    \la{acti}
    \ee
   where $H_0= 1 + {Q^{(2)}_0\over r^5}, \ \ r^2 = X_i X_i $.
   Equivalently,
    \be
    I_{2+0}  = -T_2 V_2  \F \int dt  \bigg(1 +
      H_0^{-1} \bigg[   \sqrt { (1-  H_0 v^2)( 1 +  H_0\F^{-2} ) }
      - 1 \bigg]   \bigg) \ ,
    \la{arr}
    \ee
    where   according to \rf{inte}  $T_2 V_2  \F =T_0$.
     The  action \rf{acti}  (up to the CS term) has a remarkable symmetry between the $v$-
      and $\F$-
   dependence
     which is,  of course,   a  consequence of the relation
      between the BI action and the
      D-brane action
      via T-duality  ($v$ is the counterpart of the
       electric field \ci{bachas}).

 The case of zero $\F$ corresponds to the  0-brane -- 2-brane interaction; for non-zero
 separation
 this is not a BPS configuration  and there is an attractive static
  potential \ci{pol,lif2,ts3}. On the other hand, for large $\F$ and at
   large distances (when
$H_0-1$ is small)    the leading term in the expansion  in $\F\inv$ cancels
 out just
as in
 the  case  of 0-brane -- 0-brane  scattering (which is a BPS  configuration
for zero velocity).
The  2-brane with
large  $\F$ looks  like a large number of  0-branes  smeared over a torus.
  Its superposition with another
0-brane  is approximately BPS
  as their interaction   is dominated by
  the interaction of a large number of  0-branes on 2-brane
  with the   0-brane.   As a result,
   the static potential vanishes  as  $O(\F^{-3})$
   (see  below).

 Separating the leading  long-distance   interaction term
 we get,  for general $\F$,
      \be
 I_{2+0} =  \int dt  \bigg[ -  T_{2+0}   \sqrt { 1-   v^2}  -  \V (v,r)   \bigg] \ , \la{tec}
 \ee
 where
 \be
   T_{2+0} =  T_2 V_2 ({1 + {n^2_0\ov n^2_2} })^{1/2}
   = T_0 (1 + {n^2_2\ov n^2_0} )^{1/2}   = g\inv (2\pi T)^{1/2}
   \sqrt {n^2_0 +n^2_2}   \ ,  \la{terr}
   \ee
 and
 $$
 \V  = -  {1 \ov 2 r^5} Q^{(2)}_0  T_2 V_2
   \sqrt { 1+   \F^2}\sqrt { 1-   v^2}	 \bigg( { 1\ov  \sqrt { 1-   v^2} } -  {\F \ov \sqrt {1 +
   \F^2} } \bigg)^2
 + O({1\ov r^{10}})
 $$
 \be
 =  - { 3\ov  4 r^5} N_0 \sqrt {n^2_0 +n^2_2} \sqrt { 1-   v^2}
 \bigg( { 1\ov  \sqrt { 1-   v^2}} -  {n_0   \ov \sqrt {n_0^2  +
   {n^2_2} } } \bigg)^2  + O({1\ov r^{10}})  \ .
 \la{poot}
 \ee
 Equivalent expression  was found from the open string theory
 representation (annulus diagram) in \ci{lif2}.
 The potential $\V_{string}$ in \ci{lif2} is related to the static
 gauge potential $\V$ by $
\V_{string} =  { 1 \ov  \sqrt { 1-   v^2}}\ \V  , $
 so that  $I_{2+0} =
  - \int dt  \sqrt { 1-   v^2} \big( T_{2+0} +  \V_{string}\big)$.

 To compare  with the M(atrix) theory result
  we need to expand in $v$ and to
 assume that  the  number of 0-branes is large,
   $n_0\equiv N\to \infty$ (in the  M(atrix) model \ci{bfss,grt}
  the boosted M2-brane  represented by  the
     constant  gauge field strength solution  corresponding to the magnetic
      flux on the dual torus).
    This gives
    $$
  \V_{v\to 0, n_0 \to \infty}   =
   -  { 3\ov   16  r^5} N_0  \sqrt {n_0^2 + {n^2_2}}  \sqrt { 1-   v^2}
    \bigg[ \bigg( {n^2_2 \ov  n_0^2}   + v^2 \bigg)^2  + ...\bigg]
     +  O({1\ov r^{10}})
    $$
    \be
      =
   -  { 3\ov   16  r^5} N_0 n_0    \bigg[ {n^4_2 \ov  n_0^4}
   + 2v^2 {n_2^2 \ov  n^2_0} 
    +  v^4    + O({n_2^6 \ov  n^6_0}, {n_2^4 \ov  n^4_0}v^2,v^6) \bigg]
      +  O({1\ov r^{10}}) \ ,
   \la{appo}
   \ee
   where $n^2_2\ov n^2_0$ and $v^2$  are assumed to be of the same
    order (then $\V$ and $\V_{string}$ have the
    same first three terms in their $v\to 0, n_0 \to \infty$
    expansion). 
   This  potential was shown to be
    in  agreement with  the long-distance   M(atrix) theory
    result in \ci{lifmat}.
    
   The exact expression \rf{poot}
    was also obtained  \ci{ballar}
    from supergravity  by a different method  (based on
   solving the  Hamilton-Jacobi
    equation for a 0-brane propagating  in  the $2+0$  background).
   The  relation between the phase shift and the potential
   \be
   \delta (r,v) = - \int^\infty_0  d\t \ \V(r(\t),v) \ , \ \ \ \
   r^2(\t) = r^2 + {v^2 } \t^2\ ,
   \la{paas}
   \ee
   gives, in the  case of  $\V = f(v)  r^{-5} ,$ \
  $
   \delta = - {2r   \ov 3 v}    \V  =
   - {2r   \ov 3 v}  \sqrt { 1- v^2}   \V_{string}$, 
which  for $\V$ in  \rf{poot} gives the same result  as in  \ci{ballar}.
Thus the  full $(v,\F)$-dependent phase shift  can be found
from the $D=3$ BI action for the $2+0$ brane probe.\foot{It is not clear if
 this exact expression can be derived  from the  one-loop  effective action
 of the
SYM theory   modified by higher-order terms
since the short-distance and long-distance  forms  of the string
phase shift  agree only for two leading powers in the  small $v$ expansion.
We are grateful to G. Lifschytz for this remark.}
 The advantage of the probe method is its universality, simplicity
 and direct relation to D-brane  action. This
 will be  illustrated further on the examples discussed below.

%%%%%%%%%%%%%%%%%%%%%%%%%%%%%%%%%%%%%%%%%%%%%%%%%%%%%%
\subsection{ $(2+0)$ - $(4\pa 0) $ interaction }
%%%%%%%%%%%%%%%%%%%%%%%%%%%%%%%%%%%%%%%%%%%
   The probe method  described above  makes it easy to
    find
     the classical long-distance
     interaction potential  between  boosted M2-brane and  longitudinally
   boosted M5-brane, i.e. between $2+0$ and $4\pa 0$  D-brane bound states.
   One    may  consider either
   (i)  a $2+0$ probe in $4\pa 0$ background, or (ii)  a
   $4\pa 0 $ probe in $2+ 0$ background.
   Here we shall follow the first approach as it is closely
   related to the  discussion of the previous section.
   The use of $4\pa 0$ as a probe will be discussed in section 2.4  below.

  The starting point is again the action for the $(2+0)$-brane probe, i.e.
    \rf{act}
  with  non-vanishing magnetic background \rf{inte}.
  We shall choose to orient 2-brane orthogonally to the 4-brane
  (the cases of other orientations are discussed in a similar way).
  We shall assume that the 4-brane is wrapped over a 4-torus in  directions
  $1,2,3,4$ and the 2-brane is wrapped over a 2-torus in  directions
  $5,6$.
   The orthogonally  oriented and separated  2-brane and 4-brane is not
   a BPS configuration, i.e. there will be a non-vanishing static potential.
   However, in the limit of  large 0-brane content  in  the $2+0$ state,
   i.e. large  flux $\F$, the static potential will vanish  as $O(\F\inv)$.
   The reason is that for large $\F$ the $2+0$ brane  will behave essentially
   as a collection of 0-branes on $T^2$ and thus will form
   a BPS configuration when superposed with $4\pa 0$ (a static
   configuration of a 4-brane and a 0-brane is BPS).
   This configuration will be approximately supersymmetric (as in the
   cases discussed in \ci{lifmat,lif3}),
   and thus the leading terms in the
    potential computed in string theory  will  have
    the same short-distance (open-string determined) and long-distance
   (closed-string determined) behaviours,
   allowing one to  expect that the long-distance  potential  computed
   using  closed string effective action  (supergravity)
   will be equivalent to  the one-loop potential computed using
   SYM or M(atrix) theory.
   This  indeed  is  what we  will demonstrate  below.
   The novel element compared to the discussions in
   \ci{ab,lifmat,lif3} and the previous subsection
   is that here one of the two objects ($4\pa 0$)
   will have only 1/4 of maximal supersymmetry.

    The $4\pa 0$  type  IIA  supergravity background
    smeared in  the two  directions $(5,6)$ parallel to the 2-brane
    probe   is ($i=7,8,9$)  \ci{ts2}
   \be
   ds^2_{10} = (H_0H_4)^{1/2}  [ - (H_0 H_4)^{-1} dt^2 +
  H^{-1}_4 (dy_1^2 + ...+ dy_4^2)  +  dy_5^2 + dy_6^2 +
    dx_i dx_i ] \ ,
    \la{mee}
    \ee
$$ e^{\phi} = H_0^{3/4}H_4^{-1/4}  \ ,  \ \  \ \ \  \ \
C_0 = H_0^{-1}-1  \ ,
\ \ \ \ \ \   H_0= 1 + {Q^{(6)}_0\over r} \ , \ \ \ \ \ \
H_4= 1 + {Q^{(2)}_4\over r} \ ,   $$
where $Q^{(n)}_p$ are given by \rf{sme}.
The magnetic $C_{mnk}$ background ($dC_3= *dH_4\wedge dy_5\wedge dy_6$)
will not be relevant   in the present case
 ($C_3$ background will be  contributing   in the case of parallel
  4-branes discussed in section 2.4).

 Ignoring the dependence on spatial derivatives of $X_i$, 
 we find from \rf{act}
the following  $2+0$ probe action   (cf. \rf{acti},\rf{arr})
      $$
 I_{2+0}  = -T_2  \int d^3 x
  \bigg[ H_0^{-1}  \sqrt { (1-  H_0H_4  v^2)( \F^2 + H_0H_4  ) }
    - (H_0^{-1} -1)\F   \bigg]
    $$ \be
    = -T_2 V_2 \F  \int d t
  \bigg( 1 + H_0^{-1} \bigg[  \sqrt { (1-  H_0H_4  v^2)(1  + H_0H_4 \F^{-2}  ) }
    - 1\bigg]   \bigg)  \ .
     \la{accq}
   \ee
   where it is  assumed that $\F$ has a constant magnetic background
   given by \rf{inte}.

   As expected, this  action  takes the same form as \rf{acti} when
   $H_4=1$.
   For zero $\F$ it gives the  interaction  potential (with non-vanishing
   static part)   between $4\pa 0$ and
    orthogonal 2-brane \ci{ts3}. For large $\F$  (and $v=0$) this
    configuration becomes approximately BPS,
     and the leading term in the potential  is proportional to $v^2$ as
     in the case of $0$-brane -- $4\pa 0$ scattering discussed
     in the  next section.

  For general $\F$ the
   leading long-distance interaction  potential  $\V$ in  \rf{accq},\rf{tec} is thus
$$
 \V =  -  {1 \ov 2 r}   T_2 V_2
   \sqrt { 1+   \F^2} \sqrt { 1-   v^2}
   	 \bigg[ Q^{(6)}_0 \bigg({ 1\ov  \sqrt { 1-   v^2} } -  {\F \ov \sqrt {1 +
   \F^2} } \bigg)^2
   $$ \be   + \
    Q^{(2)}_4  \bigg( { v^2 \ov { 1-   v^2} }   - {1 \ov {1 + \F^2} } \bigg) \bigg]
 + O({1\ov r^{2}}) \  .
 \la{smt}
 \ee
 This  expression can be simplified  by assuming that the 2-torus and 4-torus
  are self-dual \rf{self} so that
  according to
 \rf{sme}   \
 $
 Q^{(6)}_0 = N_0 g (8\pi T)^{-1/2}, \ \
  Q^{(2)}_4 = N_4 g (8\pi T)^{-1/2} $\  ($\om_0=2, \ \om_2= 4\pi$).
 $N_0$ and $N_4$ are the charges of the source configuration.
 Then
 \be
 \V  = -  {1 \ov 4 r}
   \sqrt { n^2_2 +   n_0^2 } \sqrt { 1-   v^2}
   	 \bigg[ N_0  \bigg( { 1\ov  \sqrt { 1-   v^2} } -
   	 {n_0 \ov \sqrt {n^2_2 +
   n^2_0 } }\bigg)^2   +  N_4  \bigg( { v^2 \ov { 1-   v^2} }
   - {n^2_2 \ov {n^2_2 + n_0^2 } } \bigg)
   \bigg]
 + O({1\ov r^{2}})   .
 \la{voot}
 \ee
 The leading terms in    $n_0 \gg n_2$ ($\F \to \infty$)  expansion
 of  static part of this
   potential   are
 $$
 \V   = -  {1 \ov 4 r} \sqrt { n^2_2 +   n_0^2 }
   	 \bigg[ N_0  \bigg( 1 -  {n_0 \ov \sqrt {n^2_2 +   n^2_0 } }\bigg)^2
   	    -  N_4 {n^2_2 \ov {n^2_2 + n_0^2 } }
+  O(v^2)  \bigg]
 + O({1\ov r^{2}})   \
 $$
 $$
 = {1 \ov 4 r}  {n_0 n_2 \ov \sqrt{n^2_0  + {n^2_2 }} }  
    \bigg[  {n_2 \ov n_0} N_4 
 -   {n^3_2 \ov 4 n^3_0} N_0   +  O({1 \ov  n^5_0}, v^2)  \bigg]
 +  O({1\ov r^{2}} ) \ 
 $$
 \be 
 = {1 \ov 4 r}    
    \bigg[  {n^2_2 \ov n_0} N_4 
 -   {n^4_2 \ov 4 n^3_0} ( N_0 + 2 N_4)   +  O({1 \ov  n^5_0}, v^2)  \bigg]
 +  O({1\ov r^{2}} ) \ . 
 \la{vot}
 \ee
 We shall reproduce this  static 
 expression\footnote{The velocity-dependent terms in this potential 
are, of course, consistent with  the $0$--$(2+0)$ result 
 \rf{appo} in the limit when $N_4=0$:\ \ 
$\V= {n_0 \ov 4 r}    
    \big[  {n^2_2 \ov n^2_0} N_4 
 -    \fourth ({n^2_2 \ov  n^2_0} + v^2)^2  N_0 + ... \big]
 +  O({1\ov r^{2}} ) . $}   
for $N_0 \gg N_4$
 as  the  long-distance
 potential between the
 M2-brane and longitudinal 5-brane in  M(atrix)   theory
 in section 3.3.

%%%%%%%%%%%%%%%%%%%%%%%%%%%%%%%%%%%%%%%%%%%%%%%%%%%%%%
\subsection{ $0$ - $(4\pa 0) $ interaction }
%%%%%%%%%%%%%%%%%%%%%%%%%%%%%%%%%%%%%%%%%%%
Let us now consider the closely related case of  0-brane
scattering off $4\pa 0$ bound state.
This may be viewed as a  special case
 of the $(2+0) - (4\pa 0)$
scattering in the limit  when the magnetic field on 2-brane
(i.e. the 0-brane charge)
becomes large. The resulting expression for the potential will be very
similar to the $\F\to \infty$ limit of
 \rf{smt}; the only difference is that here we will not assume that
two extra transverse directions are compactified on 2-torus.

  We shall consider a 0-brane probe  in the
$4\pa 0$ background, but the same result is found by
studying the $4\pa 0$ probe moving in the  0-brane background (which  will be
 a special
case of the discussion  in section 2.4).
The  action for a 0-brane probe moving orthogonally to
  the $4\pa 0$  source wrapped over $T^4$  is  found to be \ci{ts3}
 \be
 I_0 = -T_0 \int dt  \bigg(1 +  H_0^{-1}  \sqrt {1- H_0 H_4 v^2  }
    -   H_0^{-1}  \bigg) =
    \int dt  \bigg( -T_0 \sqrt {1-  v^2  } - \V  \bigg)  \ ,
 \la{zerr}
 \ee
 \be H_0= 1 + {Q^{(4)}_0\over r^3} \ , \ \ \ \ \ \  H_4= 1 + {Q_4\over r^3}
 \ , \ \ \ \  Q^{(4)}_0 = N_0 N_4\inv  Q_4  \ .
 \la{rele}
 \ee
 It has, indeed, the same form as the large $\F$ limit of \rf{accq}
 with $T_2V_2 \F=T_0$  held fixed.

The leading term in the long-distance potential is
    \be
 \V =  -  {1 \ov 4 r^3}  n_0   T\inv
    {1 \ov \sqrt { 1 -   v^2} }
     \bigg[  N_0 \big({  \sqrt { 1-   v^2} }  -  1  \big)^2
    +  N_4  { v^2  }  \bigg] + O({1\ov r^{6}})\ ,
 \la{smti}
 \ee
 i.e. is  the same as
 the $n_0 \gg n_2$ limit of \rf{voot} with $ r\inv  \to   T\inv r^{-3}$.
 Its low-velocity expansion is
 $$\V = -  {1\over 4r^3 } n_0  T\inv \bigg[ {1 \ov \sqrt { 1 -   v^2} }
 \bigg( N_4 v^2   + {1\over 4} N_0 v^4   \bigg)+  O(v^6)
\bigg]  + O({1\ov r^{6}}) \  $$
 \be
=  -  {1\over 4r^3 } n_0  T\inv
 \bigg[ N_4  v^2  + {1\over 4} (N_0 + 2 N_4) v^4 +  O(v^6)  \bigg]
 + O({1\ov r^{6}})
\  . \la{new}
\ee
 The resulting  phase shift $\d$
 \rf{paas}
 in the  present case of $\V = f(v)/r^3$  is  given by
 \be
 \delta(r,v) = - {2r v\inv } \V   =
  {1\over 2r^2 } n_0  T\inv
 \bigg[ N_4 v   + {1\over 4} (N_0 + 2 N_4) v^3  \bigg] +  O(v^5) \ .
\la{dee}
 \ee
 As we shall show in section 3.2, this expression is again in agreement
 with the corresponding  M(atrix)  theory calculation (for $N_0 \gg N_4$).

 The  potential $\V_{string}$  equivalent to \rf{smti}
 was found in \ci{lif2} by considering the scattering of a 0-brane off the
  1/2 supersymmetric $4+2+0$  non-marginal bound state  in the open string
 theory representation. The $4+2+0$  configuration
 was described as a 4-brane with two  constant magnetic  fluxes
  $\F_{12}$ and $ \F_{34}$, implying the presence of
  the  0-brane charge
 $N_0 \sim \F_{12} \F_{34}$ and the two 2-brane charges
 ($\sim \F_{12}$ and $\sim \F_{34}$)
 in the two orthogonal 2-tori.
  Such  4-brane
  becomes similar to a $2+0$ brane  in the limit of large  $\F_{12}$
  (or large $\F_{34}$).  When both $\F_{12}$ and $\F_{34}$
  are large  and equal the coupling to $C_3$
   (i.e. the 2-brane charge content)
  is suppressed  and  $4+2+0$ configuration
  becomes   essentially similar to the  $4\pa 0$ bound state  with
  large 0-brane content.
  The 2-brane charges also do not contribute in the case of
  0-brane  scattering off a  4-brane
  with  magnetic fluxes. For {\it equal} fluxes, 
  i.e.  
  a  self-dual magnetic background, 
  the static force  
  vanishes, 
  and the resulting  potential is the {\it same}
   as in the case of the
  0-brane scattering off the $4\pa 0$ marginal bound state
  (see also  the discussion in
  section 2.4).
  This explains the agreement between \rf{smti} and the expression found 
  in \ci{lif2}.

  The two leading terms ($\sim N_4v + \fourth N_0 v^3, \ N_0 \gg N_4)$  in the phase shift 
  \rf{dee} were found in  \ci{lif3} to be in agreement with the
  corresponding  matrix model
  calculation -- the  scattering of a  0-brane off  the
  4-brane with { large}
  and  equal magnetic  fluxes ($\sim 1/\sqrt{N_0}$)
   in the two orthogonal planes.
  The  result of our  matrix model calculation in section 3.2
  for the potential between a  0-brane and a $4\pa 0$  bound state
  will  thus be essentially the same  as in  the case of
  the 0-brane -- $4+2+0$ brane scattering  considered in  \ci{lif3}.

%%%%%%%%%%%%%%%%%%%%%%%%%%%%%%%%%%%%%%%%%%%%%%%%%%%%%%
\subsection{$(4\pa 0) $ as a classical probe: $(4\pa 0)$ - $(4\pa 0) $
interaction}
%%%%%%%%%%%%%%%%%%%%%%%%%%%%%%%%%%%%%%%%%%%%%%%%%%%%%%%%%%%%%

To study the  interaction of two parallel longitudinal 5-branes,
i.e.  two marginal $4\pa 0$ D-brane bound states,
we shall consider the $4\pa 0$ system as a probe in the background
produced  by another $4\pa 0$ as a source.
The 4-branes will be assumed to be wrapped over  $T^4$.
The action for the classical   \fo   probe will be taken
 to be the standard
D4-brane action with an extra (constant, abelian, magnetic)
{\it self-dual}  world-volume gauge field background.
The presence of a non-trivial  gauge field
is necessary in order to induce  the 0-brane charge  on  4-brane \ci{doug},
and self-duality is crucial  for  correspondence with the
 properties of a  marginal $4\pa 0$  configuration (see below).

The  action for a 4-brane
    in a  curved background is
    ($ m,n=0,..,4; \ i,j=5,...,9$)
 $$
 I_4 = - T_4\int d^5 x  \bigg[  e^{-\phi} \sqrt {-\det\big( G_{mn } +
 G_{ij} \del_m X^i \del_n X^j +  \F_{mn} \big)} $$
 \be
    -\   {1\over 5!}  \e^{mnkpq} C_{mnkpq}   -   {1\over 12}  \e^{mnkpq} C_{mnk}  \F_{pq}  -
    {1\over 4}  \e^{mnkpq} C_m \F_{nk} \F_{pq} \bigg] \ ,
    \la{actf}
    \ee
    where $dC_5 = *dC_3+...$ (dots stand for $C_3 \wedge dB_2 + *(B_2 \wedge
    dC_1)$  terms which will not be relevant here). 
      As in \rf{act},   we used the static gauge and
    assumed that the metric is `diagonal'.
     We shall choose $\F_{mn}= T\inv F_{mn} $ \ ($B_{mn}=0$)    to  have only spatial ($a,b=1,2,3,4$)
     components   and to be self-dual   (cf. \rf{inte})
\be
\F_{ab} = *\F_{ab}\equiv  \hal e_{abcd} \F_{cd}
 \ , \ \ \ \    \fourth T_4 V_4  \F_{ab} \F_{ab} =  T_0 \ , \ \ \ \
{\rm i.e.} \ \ \ \ \fourth  \F_{ab} \F_{ab} = {n_0 \ov n_4} \ ,
\la{fie}
\ee
where in the last relation we assumed
that the 4-torus has  self-dual  value of the volume \rf{self}.
Here the contractions of repeated $a,b$ indices are with flat 4-metric.
The explicit form of
     such  abelian instanton configurations on
      $T^4$ (see \ci{vbaal,hashi})
     will not be important here.

In general, introducing a  constant  magnetic
flux on a 4-brane wrapped over $T^4$  induces
the 0-brane charge $n_0 \sim \F_{ab} *\F_{ab}$
and the 2-brane charges $n_2 \sim \int_2  \F $ in the  corresponding
2-cycles. If $\F_{ab}$ is put in a block-diagonal
form with   eigenvalues $f_1=\F_{12}$ and $f_2= \F_{34}$
then $N_0 \sim f_1 f_2$ and the  2-brane charges in the two
orthogonal (12) and (34)  2-tori  are $n_2 \sim f_1, \ n_2'\sim f_2$.
The 4-brane with such flux \ci{lif2} thus represents the 1/2 supersymmetric
non-marginal bound state  $4+2+0$ (or
 `$4+ 2\perp 2+0$').
 The corresponding tension  as determined by the
  BI term in
the flat-space action  \rf{actf} is
 $$T_{4+ 2\perp 2+0} = T_4 \sqrt{(1+ f^2_1) (1+f^2_2)}  =
  \bar T_4 \sqrt{n^2_4 + n^2_2 + n'^2_2 + n_0^2} \ . $$
 It may seem  that one cannot use such 4-brane with a flux
 to represent  the  marginal 1/4 supersymmetric $4\pa 0$
 bound state. Note, however,
 that  when $\F_{ab}$ is {\it self-dual}, i.e. when 
 $f_1=f_2=\F, \ \F^2 = \fourth \F_{ab} \F_{ab} $, then $n_2=n_2'$ and thus
  (see also below)
 $$ (T_{4+ 2\perp 2+0})_{\rm self-dual} \  =\  T_4(1+ \F^2)   =
  \bar T_4 (n_4 + n_0) =T_{4+0}  \ ,  $$
 i.e. the tension is the same as  for the  marginal $4\pa 0$ bound state.

 There are circumstances under  which  the 4-brane
 with a constant self-dual magnetic flux can, indeed,
 be used  as a probe representing the $4\pa 0$  marginal
 bound state. For example,
 if the 4-brane  probe   is put in a background which does not couple to
the  2-brane charges (e.g., having $C_3=0$)
then the motion of the probe   will be determined only by the
4-brane and 0-brane couplings.
As we shall demonstrate below,  in this case interpreting
such 4-brane as a $4\pa 0$ probe
leads to consistent results.
Another  situation when the contribution of 2-brane charges will be
suppressed is the  large flux limit,
in which $n_0 \gg n_2,n_2'$ (in this case the $4+2\perp 2 +0$ system
is dominated by the 0-brane charge).\foot{Though this is not
suitable for a classical probe  picture,
 it is    possible also   to consider (as in
 \ci{wit3,doug})  a non-abelian generalisation
 of \rf{actf} and to assume that a non-abelian
   self-dual background is such that
 $\int d^4 x  {\ \rm tr} ( \F_{ab} *\F_{ab}) \not=0$ while
 ${\rm tr}\F_{ab}=0$   so that  only the 0-brane charge is present. 
 This corresponds to the M(atrix) model description of the $4\pa 0$ system.}

Let us  now   study  the transverse motion of the \fo probe
represented as a   4-brane with a self-dual $\F_{ab}$ gauge field 
in the 1/4 supersymmetric  background   corresponding to the
marginal \fo bound state
  (i.e. in the `unsmeared' version of \rf{mee}). If we  ignore   the dependence  of the probe action on
  the spatial  derivatives of $X_i$,
     then the 5-dimensional determinant in \rf{actf}
     factorises  into the product of the  $(00)$-element  and a 4-dimensional
     determinant. Using the  curved-space generalisation
     of the standard    relation  (relevant for the $D=4$ BI action)
     \be \det_4 (\d_{ab} +  \F_{ab})=
     1  + \hal   \F_{ab} \F_{ab}  +
     {1\ov 16}  (\F_{ab} *\F_{ab})^2
      =
     (1 + {1\ov 4} \F_{ab} *\F_{ab})^2 + {1\ov 4} (\F_{ab} - *\F_{ab})^2
     \ , \la{rer} \ee
     we conclude that in the case of a  {\it self-dual}  field strength
   the $\F_{ab}$-dependent part of the expression under   the square root in  \rf{actf} is
   a total square. The
   action \rf{actf} thus  becomes {\it quadratic}  in $\F$, and
      takes the following  remarkably simple  form  (cf. \rf{zerr})
 $$
 I_{4\pa 0} = - T_4   \int d^5 x  \bigg[
 \bigg( 1 +{1\over 4} \F_{ab} \F_{ab}H_4 H_0^{-1}  \bigg)
 H_4^{-1}\  \sqrt {1-  H_0H_4  v^2} $$
  \be
  - (H_4^{-1}-1)  - {1\over 4}\F_{ab} \F_{ab} (H_0^{-1} -1) \bigg] \  ,
    \la{acci}
    \ee
    or
    $$
   I_{4\pa 0}  =- T_4   \int d^5 x  \bigg[ 1 +  H_4^{-1} \big(  \sqrt {1-  H_0H_4  v^2}
    -1\big)
    +  {1\over 4}\F_{ab} \F_{ab} \big(1 +  H_0^{-1}[\sqrt {1-  H_0H_4  v^2}
    -1]\big)
\bigg] \ . $$
    Here $H_0, H_4$ are the same as in \rf{rele} and
    the last two terms in \rf{acci} came from the  CS terms
    in \rf{actf}.
    Note that the
    $C_3\wedge \F$ term in the action \rf{actf}
    vanishes  as
    both $C_3$ and $\F$ have only magnetic backgrounds, i.e. the  
    2-brane charges  on the 4-brane 
    induced by the presence of the magnetic
    $\F_{ab}$-fluxes  do not contribute in the present case

    Using \rf{fie} and integrating over the spatial world-volume
    coordinates (assuming that $X_i=X_i(t)$)
     \rf{acci} can be written also as
   $$
 I_{4\pa 0}    =-  \bar T_0  \int dt
  \bigg( n_4 + n_0  +  n_4  H_4^{-1} [  \sqrt {1-  H_0 H_4  v^2}
    -1]
    +   n_0   H_0^{-1}[\sqrt {1-  H_0 H_4  v^2} -1]
\bigg) $$
\be
  =- \int dt \bigg( T_{4+0} \sqrt {1 - v^2} - \V \bigg) \ ,  \la{obvi}
\ee
where $\bar T_0= g\inv (2\pi T)^{1/2}$ and 
 $T_{4+0}= \bar T_0 (n_0 + n_4)$ 
is the  mass of the  $4\pa 0$ `particle'
which,  indeed,  has the
value  expected  for a marginal $4\pa 0$  bound
 state.
 The action has the  obvious $0 \leftrightarrow 4$ invariance (implied by
 T-duality)
 and reduces to \rf{zerr} for $n_4=0$.

 The static part of the potential $\V$ vanishes as it should  for the parallel
 $(4\pa 0)\  \pa\  (4\pa 0)$ configuration which  is  BPS for
 $v=0$.
 The leading long-distance term in  $\V$  is  a  generalisation of
 \rf{smti}
    \be
 \V =  -  {1 \ov 4 r^3}     T\inv
    {1 \ov \sqrt { 1 -   v^2} }
     \bigg[ ( n_0 N_4 + n_4 N_0)  { v^2  }
     + (n_0 N_0  + n_4 N_4) \big({  \sqrt { 1-   v^2} }  -  1  \big)^2
     \bigg]
    + O({1\ov r^{6}})\ .
 \la{mti}
 \ee   
 From    the  low-velocity expansion of $\V$
 $$
 \V=-  {1 \ov 4 r^3}     T\inv
    {1 \ov \sqrt { 1 -   v^2} }
     \bigg[ ( n_0 N_4 + n_4 N_0)  { v^2  } +
         \fourth  (n_0 N_0  + n_4 N_4) v^4  + O(v^6)    \bigg]
    + O({1\ov r^{6}})\ 
    $$
    \be
=   -  {1 \ov 4 r^3}     T\inv
     \bigg[ \big( n_0 N_4 + n_4 N_0\big)  { v^2  } +
         \fourth  \big(n_0 N_0  + n_4 N_4 + 2n_0 N_4 + 2n_4 N_0
           \big) v^4     \bigg] + ... \   ,  
    \la{lowe}
    \ee
 we find the corresponding leading terms in the phase shift (cf. \rf{dee})
 \be
 \delta = - {2rv\inv }  \V  =
  {1\over 2r^2 }  T\inv
 \bigg[( n_0 N_4 + n_4 N_0) v   + {1\over 4}(n_0 N_0  +  n_4 N_4
  + 2n_0 N_4  +  2n_4 N_0 ) v^3  \bigg] .
\la{deed}
 \ee
 This   will be  checked against the M(atrix) model
 computation of  scattering of longitudinal 5-branes in section 3.4.

 The expressions \rf{mti},\rf{deed} have  a  natural structure reflecting
  the fact that $4\pa 0$ is a   marginal bound state --
 its constituents interact with external states  almost independently
 (to  leading orders in  long-distance and velocity expansions).
 Indeed, in the case of the 0--0 or 4--4 scattering the force
  starts
 with a  $v^4$ term, while in the 4--0 scattering it  contains
 already the $v^2$ term.\foot{Note that the potential for parallel
 $4+2 +0$ non-marginal bound states  starts with a $v^4$ term
 \ci{lif3}.}
  The manifest  `probe $\leftrightarrow$ source' symmetry
 of \rf{mti}
 confirms the consistency of our interpretation
 of the 4-brane action with  self-dual $\F_{ab}$ field
   in $4\pa 0$ background as
 describing the $(4\pa 0)$ -- $(4\pa 0)$ interaction.

 The phase shift \rf{deed} should have the same form
as  for the  scattering
 of wrapped fundamental strings   in the
  (momentum, winding) BPS states
 (see \ci{polchi,calkhu}),
 since by U-duality they  are related to the  $4\pa 0$ bound states.
 It should be  possible also to reproduce \rf{deed}
 by a classical  supergravity calculation of scattering of  the
  corresponding
  extremal 1/4 supersymmetric
 black holes with two charges   following  the method used
 in the   single-charge  1/2
 supersymmetric case  in \ci{shira}.
 The probe method used here   has the  obvious
 advantage of simplicity.

%%%%%%%%%%%%%%%%%%%%%%%%%%%%%%%%%%%%%%%%%%%%%%%%%
\newsection{M(atrix) theory  (SYM)  description of  longitudinal 5-brane
interactions}
%%%%%%%%%%%%%%%%%%%%%%%%%%%%%%%%%%%%%%%%%%%%%%%%%%%%%%%%%%%%%%%
The M(atrix) theory Lagrangian  \ci{bfss} is the  10-dimensional 
  $U(N)$
SYM  Lagrangian reduced
to 0+1 dimensions (in this section we shall set $T\inv = 2\pi \a'=1$)
\be
L=\frac{1}{2g_s}\tr \bigg( D_t X_i D_t X_i + 2\te^{T}D_{t}\te
-\frac{1}{2}\lsb X_i,X_j\rsb^2-2\te^{T}\gamma_i \lsb \te ,X_i\rsb \bigg) \,,
\label{lagran}
\ee
where $X_i$ and $\te$ are  bosonic and fermionic hermitian
$N\times N$ matrices  and
$
D_t X=\de_tX-i\lsb A_0,X\rsb.
$
Below we shall demonstrate that the leading-order terms
in the long-distance, low-velocity potentials  between BPS bound states
with 1/2 and 1/4  of supersymmetry
computed  in the previous
sections  using classical closed string effective field theory
methods  are indeed   reproduced  by the  corresponding 1-loop
 SYM  computations.
%%%%%%%%%%%%%%%%%%%%%%%%%%%%%%%%%%%%%%%%%%%%%%%
\subsection{One-loop  effective action in a general
background\label{effaction}}
%%%%%%%%%%%%%%%%%%%%%%%%%%%%%%%%%%%%%%%
We shall start with   a  calculation
of the one-loop effective action in the theory
\rf{lagran} for the  relevant class of    backgrounds. Our expressions
generalise  those of \ci{ab,lifmat,lif3}.
Let us consider the following  background gauge field
\be
\bar{{\cal A}}_{\mu}=\left( \bar{A}_0=0,\bar{X}_1,\ldots
,\bar{X}_8,\bar{X}_9\right) \,,
\label{backgr}
\ee
where the time-independent components
\be
\bar{X}_i=\LB
\begin{array}{cc}
 \bar{X}_i^{(1)} & 0 \\
 0 & \bar{X}_i^{(2)} \\
 \end{array}
 \RB ,~~~~~~i=1,\ldots,7
\ee
 correspond to the coordinates of the two
 BPS objects,
\be
\bar{X}_8=\LB
\begin{array}{cc}
 b & 0 \\
 0 & 0 \\
 \end{array} \RB
\ee
represents  their separation $b$,   and
\be
\bar{X}_9=\LB
\begin{array}{cc}
 vt & 0 \\
 0 & 0 \\
 \end{array} \RB
\ee
describes  their relative motion. The   calculation
of the  SYM one-loop  effective
action in the background \rf{backgr} is  described   in
Appendix~A. Let us define the commutators
\be \xijo = [\xo_i,\xo_j] \ , \ \ \ \ \   \xijt = [\xt_i,\xt_j]\ ,
\ee
and the operators
\be H = \LB \xo_i \otimes {\bf 1}-{\bf 1}\otimes
{\bar X}^{(2)*}_i\RB^2 \ , \ \ \ \ \
 H_{ij}=\xijo \otimes {\bf 1}+{\bf 1}\otimes
{\bar X}_{ij}^{(2)*} \ ,
\ee
where $*$ is the complex conjugation.  In this
notation, the effective action reads
\be W=-\hbox{ln~} ( BGF )  \,,
\label{effact}
\ee
where
$B$, $G$ and $F$ are the bosonic, ghost and fermionic contributions,
respectively. $B$  is given by
\be
B=\bigg(\hbox{det}\lsb \LB -\deu +H \RB
\delta_{\mu\nu}+ E_{\mu\nu}+2H_{\mu\nu}\rsb \bigg)^{-1}\,,
\ee
where $E_{\m\n}$ is the matrix with $E_{90}=-E_{09}=2v$ as
the only  non-vanishing components.
The  differential operator under   the determinant
acts on on the space of functions of   of Euclidean
 time  $\tau$  and also on the  $U(N)$ matrix index  space
 and  the Lorentz vector   space.

For $G$ we have
\be
G=\bigg[\hbox{det} ( -\deu + H)  \bigg]^{2}   \, ,
\ee
where the differential operator  acts  also in the matrix index space.
The fermionic contribution $F$ is
\be
F= \bigg[ \hbox{det} \big( -\deu
+H+\sum_{i<j}\gamma_i\gamma_jH_{ij}+ i \gamma_9v \big) \bigg]^{1/2}  \ ,
\ee
where the operator  acts in  the  matrix space and Lorentz
spinor space.

%%%%%%%%%%%%%%%%%%%%%%%%  GRAVITON-FIVEBRANE  %%%%%%%%%%%%%%%%%%%%%%%%%%%%%%%%%%
\subsection{Graviton - longitudinal 5-brane ($0$ - $4\pa 0$)
 interaction }
 %%%%%%%%%%%%%%%%%%%%%%%%%%%%%%%%%%%%%%%%%%%%%%%%%%%%%%%%%
Let us consider the scattering of  a $D=11$ graviton (i.e. a bound state of
  $0$-branes  of charge $n_0$)  off
  a longitudinal 5-brane   (i.e.
$4\pa 0$ bound state of  $4$-brane with  charge $N_4$ and
$0$-brane with charge $N_0$).
In the M(atrix) theory  language,  the longitudinal 5-brane
is described  by a configuration
corresponding to  a  $U(N_0)$
instanton   \ci{grt,bss}.
In the case of the 5-brane wrapped over a  4-torus
this is T-dual  to the  string theory description used in  section 2.4
(with  $T^4\to \td T^4$).
 The $0$-brane of
charge $n_0$ located at $x_1=\cdots =x_9=0$ is represented  by the
$\bar{X}_i=0_{n_0\times n_0},~~i=1,\cdots ,9$.
The SYM  background  describing  the scattering
of the $0$-brane off $4\pa 0$
bound state with  impact parameter $b$ and velocity $v$ is thus
\bea
\xo_a&= &i\de_a+A_a\equiv P_a\,,~~~~~a =1\,, \dots \,, 4 \,, \non
\xo_8&= &b \,, \non
\xo_9&= &vt \,, \non
\xt_i&= &0_{n_0\times n_0}\,,~~~~~~~~~~~i=1\,, \dots \,, 9 \,,
\label{GFBK}
\eea
where $A_a$ is the $U(N_0)$ gauge potential  representing  the charge
$N_4$ instanton.  The field strength
\be
G_{ab}=- {i}[P_a,P_b] = \del_a A_b - \del_b A_a -i [A_a,A_b] \, ,
\ee
will be assumed to satisfy 
\be
G_{ab} =*G_{ab} \ , \ \ \ \ \ \
{1\ov 16\pi^2   } \int_{\td T^4} d^4x {\ \rm tr}( G_{ab}G_{ab}) =  N_4 \,.
\la{value}
\ee
The bosonic, fermionic and ghost contributions to the effective action
\rf{effact} in the background \rf{GFBK} are  ($G_{\m\n}$ is non-zero
only for
$\m,\n=a,b=1,2,3,4$)
\be
W_B=\sum^\infty_{n=0}\tr\hbox{ln}\lsb \LB b_n^2+P^2 \RB
\delta_{\mu\nu}+E_{\mu\nu}+ 2iG_{\mu\nu}\rsb \,,
\label{wb1}
\ee
\be
W_G=-2\sum^\infty_{n=0}\tr\hbox{ln} \LB b_n^2 + P^2 \RB  \,,
\label{wg1}
\ee
\be
W_F=-\frac{1}{2}\sum^\infty_{n=0}\tr\hbox{ln} \bigg( b_n^2
+P^2+\frac{i}{2}\gamma_{ab}G_{ab}+\gamma_9v \bigg)  \,,
\label{wf1}
\ee
where
\be
b_n^2\equiv b^2+iv(2n+1) \,.
\ee
The traces in Eqs.~\rf{wb1}, \rf{wg1} and \rf{wf1} can be calculated
as the (large-separation)
 expansions in  powers of  $1/b^2$. The details  are
given in  Appendix~B.

When $G_{ab}$ is self-dual, the potential does not contain a static term.
 The leading  term in  the phase shift
is found to be
\be
\delta =-iW = \frac{vn_0}{32\pi^2b^2}
\int_{\td T^4 } d^4x {\ \rm tr}( G_{ab}G_{ab})  +
\frac{n_0 N_0v^3\td V_4}{32\pi^2b^2} \,,
\label{phase}
\ee
where $\td V_4$ is the volume of the   torus $\td T^4$.
Using \rf{value}
and assuming that $\td T^4$ is self-dual \rf{self}
 we get
\be
\delta = \frac{n_0 }{2b^2}\bigg( N_4 v + \fourth N_0 v^3 \bigg) \ .
\la{same}\ee
This   the same  expression as \rf{dee} ($r\to b,\ T\to 1$)
found  from classical closed string theory  calculation  in section 2.3
since 
for $N_0 \gg N_4$  the  $N_4 v^3$ term
in  \rf{dee} is subleading.

The phase shift \rf{same} is also equivalent to the  result found in \ci{lif3}
for the scattering of a 0-brane off a  1/2 supersymmetric
non-marginal  bound state $4+ 2+0$ which is described in the T-dual
matrix model picture by a 4-torus  with two equal constant magnetic fluxes
$c_1=c_2= 1/\F$ in
the two orthogonal 2-cycles. This corresponds to the special case when
 $G_{ab}$
in the above expressions is taken to be abelian, constant  and self-dual with
$c_1=c_2$ as its two block-diagonal eigen-values.
The agreement is not surprising since  in the present case of 
 large $N_0$ or small 
magnetic fluxes the contributions of 2-brane charges to the scattering are
suppressed  so that $4+ 2 +0$ configuration behaves  essentially as
$4\pa 0$ one.

%%%%%%%%%%%%%%%%%%%%%%%%%%%%%  MEMBRANE-FIVEBRANE  %%%%%%%%%%%%%%%%%%%%%
\subsection{Membrane -- longitudinal 5-brane ($(2+0)$ - $(4\pa 0)$)
 interaction}
%%%%%%%%%%%%%%%%%%%%%%%%%%%%%%%%%%%%%%%%%%%%%%%%%%%%%%%%%
Let us now compute  the  static potential between
  the transversely
oriented membrane and longitudinal 5-brane.
In the M(atrix) theory approach, the   membrane (with unit  charge  and
large $p_{11}$)
wrapped over a 2-torus
can be described \ci{nic,bfss,grt}  by the matrices
$q$ and $p$ of size $n_0\times n_0$ ($n_0\to \infty$)
   satisfying
\be
[q,p]=i c {\bf 1}  \ , \ \ \ \  \ \ \ \
c= \frac{\td V_2}{2\pi n_0} \ .
\la{vall}
\ee
  $q$ and $p$ can be represented, e.g.,  as
covariant derivative operators   with a  constant background
 gauge field strength
on the torus.
In the T-dual description this corresponds to a   2-brane  with
magnetic flux \rf{inte}, where  $\F = 1/c,  \ n_2=1$.

The background corresponding to   the configuration
 of the $(2+0)$ bound state
 extended
in the directions $(X_5,X_6)$ and the $(4\pa 0)$ bound state extended in the
directions $(X_1,... ,X_4)$,  and  separated by a  distance $b$
from each other  in the $X_8$-direction is
\bea
\xo_a&= &i\de_a+A_a=P_a \,,~~~~~a=1\,, \dots \,, 4 \,,\non
\xo_8&= &b \,, \non
\xt_5 &= &q \,, \non
\xt_6 &= &p \, ,
\label{MFBK}
\eea
where $A_a$ is the same instanton field  as in the previous section.
Let us introduce the matrix $C_{\mu\nu}$ with  only two non-vanishing
components   $C_{56}=-C_{65}=c$.
The bosonic, fermionic and ghost contributions to the effective action
\rf{effact} in this background \rf{MFBK}  are found to be
\be
W_B=\sum_{l=0}^{\infty}
\int_{-\infty}^{\infty}
\frac{dk}{2\pi}\tr\hbox{ln}\lsb \LB b_{k,l}^2+P^2 \RB
\delta_{\mu\nu}   + 2iG_{\mu\nu} -2i C_{\m\n} \rsb \,,
\ee
\be
W_G=-2\sum_{l=0}^{\infty}
\int_{-\infty}^{\infty}
\frac{dk}{2\pi}\tr\hbox{ln} \LB b_{k,l}^2 + P^2 \RB  \,.
\ee
\be
W_F=-\frac{1}{2}\sum_{l=0}^{\infty}
\int_{-\infty}^{\infty}\frac{dk}{2\pi}\tr\hbox{ln} \bigg( b_{k,l}^2
+P^2+\frac{i}{2}\gamma_{ab}G_{ab}  + i \g_5 \g_6 c \bigg) \,,
\ee
where
\be
b_{k,l}^2\equiv b^2+k^2+c(2l+1) \,  .
\ee
The calculation of the  traces is  parallel to the one  described  in
Appendix~B, so we  shall omit the details.
Note that it is important to  include  the dependence  on $c$
 exactly before doing long-distance expansion.
 The leading term
in the long-distance expansion of  the one-loop
effective action  is found to be
\be
W= \frac{c}{64\pi^2b}\bigg[ \int_{\td T^4}  d^4x {\ \rm tr} (G_{ab} G_{ab})
-N_0c^2\td V_4\bigg] \ .
\ee
Substituting the expressions  for the  background fields from  \rf{vall},
\rf{value},
 and assuming  that  the volumes
 of  the tori have  self-dual  values \rf{self},  we finish with
 \be
W= \frac{1}{ 4 b} \bigg( { N_4 \ov n_0}   -  { N_0
 \ov 4 n^3_0}  \bigg) \ ,
\ee
which is the same (for $N_0 \gg N_4$) 
 as the  leading-order long-distance  potential \rf{vot}
 found in section 2.2 for $n_2=1$
(here $W\to \V, \ b\to r$).

%%%%%%%%%%%%%%%%%%%%%%%5-brane--5-brane %%%%%%%%%%%%%%%%%%%%%%%%%%%%%%%%%%
\subsection{Scattering  of two longitudinal 5-branes ($(4\pa 0) - (4\pa 0)$)}
%%%%%%%%%%%%%%%%%%%%%%%%%%%%%%%%%%%%%%%%%%%%%%%%%%%%%%%%%%%
The   M(atrix) theory  calculation  analogous  to the ones described  above  
for  the  cases of the graviton - 5-brane and membrane - 5-brane scattering 
can be performed  in the case of the  scattering
of  two parallel longitudinal 5-branes with 4 transverse directions wrapped over
$T^4$.
The corresponding 
 SYM background describing the scattering of  a $(4\pa 0)$ bound state
  (with charges $n_4$ and  $n_0\gg n_4$) off another
$(4\pa 0)$ bound state  (with charges $N_4$ and $N_0\gg N_4$) 
with impact parameter $b$ and velocity $v$ is
\bea
\xo_a&= &i\de_a+A_a=P_a \,,~~~~~a=1\,, \dots \,, 4 \,,\non
\xo_8&= &b \,, \non
\xo_9&= &vt \,, \non
\xt_a&= &i\de_a +A_a^{\prime}=P_a^{\prime} \,,~~~~~a=1\,, \dots \,, 4
\,.
\label{FFBK}
\eea
Here $A_a$ is a  $U(n_0)$ instanton field
 with  topological  charge $n_4$ and $A_a^{\prime}$ is a 
 $U(N_0)$ instanton field  with  charge $N_4$.
  $P_a$ and $P_a^{\prime}$ act on the  same  dual 4-torus $\td T^4$.
Let us define the  $U(n_0) \times U(N_0)$ covariant derivative
and field strength 
\be
{\cal P}_a=P_a\otimes {\bf 1} + {\bf 1}\otimes P_a^{\prime *}
\ , \ \ \ \ \ \ 
\G_{ab}
=- {i}[{\cal P}_a,{\cal P}_b] \,.
\label{fst}
\ee
The field strength~\rf{fst}  
satisfies  $ \G_{ab} = * \G_{ab}$  and 
\be
{1\ov 16\pi^2   }
\int_{\td T^4} d^4x
{\ \rm tr}( \G_{ab}\G_{ab}) =  n_0N_4 +N_0 n_4  \,  .
\label{value3}
\ee
The  contributions to the effective action
\rf{effact} in the background \rf{FFBK} 
have the same form as in \rf{wb1},\rf{wg1},\rf{wf1}
with $P_a\to \P_a,\  G_{ab} \to \G_{ab}$.
The traces  can again be calculated 
as the expansions in  powers of  $1/b^2$  as in    Appendix B.
As a result, 
the leading  term in  the  phase shift
is found to be
$$
\delta=-iW =\frac{v}{32\pi^2b^2}
\int_{\td T^4} d^4x
{\ \rm tr}( \G_{ab}\G_{ab})  +
\frac{n_0 N_0v^3\td V_4}{32\pi^2b^2} 
$$
\be 
= \ { 1 \ov 2 b^2} \bigg[ (n_0 N_4 + N_0 n_4) v  + \fourth n_0 N_0 v^3  \bigg]
\  , 
\label{phase3}
\ee
where in the last relation 
 we  have  used \rf{value3} and  assumed that $T^4$ is self-dual, 
i.e. $\td V_4= (2\pi)^2$. 
This is  in agreement with the  result  \rf{deed} of section 2.4 
since  for $n_0 \gg n_4, \ N_0 \gg N_4$ the  extra terms 
in \rf{deed}, i.e.  $\sim (n_4N_4 + 2 n_0 N_4 + 2 n_4 N_0)v^3$, 
are subleading compared to the $n_0 N_0 v^3$ term.
Thus   again the long-distance, low-velocity 
phase shift extracted from the 1-loop 
SYM  calculation agrees  with the classical closed string theory
(supergravity)  result. 

\setcounter{section}{0}
\setcounter{subsection}{0}
%%%%%%%%%%%%%%%%%%%%%%%%%%%%%%%%%%%%%%%%%%%%
\begin{center}
{\bf Acknowledgments}
\end{center}
We  would like to thank  P. van Baal, G. Lifschytz, W. Taylor
 and  K. Zarembo for useful discussions.
A.A.T. is grateful to the Institute of Theoretical Physics of
SUNY at Stony Brook
for hospitality during the completion of  this work
and  acknowledges also the support
 of PPARC and  the European
Commission TMR programme grant ERBFMRX-CT96-0045.
The work of I.C. was supported in part by CRDF grant 96-RP1-253.

%%%%%%%%%%%%%%%%%%%%%% EFFECTIVE ACTION IN GENERAL BACKGROUND %%%%%%%%%%%%%%%

\appendix{One-loop Super Yang-Mills effective action\label{onloop}}
Eq.  \rf{lagran}  can be written in the $D=10$ SYM form
\be
L=-\frac{1}{4g_s}\tr ({\cal F}_{\mu\nu}{\cal
F}^{\mu\nu})  +\hbox{fermionic term}  \,,
\ee
where
\be
{\cal A}_{\mu}=\left( A_0,X_1,\ldots
,X_8,X_9\right)
\ee
and
\be
{\cal F}_{0i}=\de_0X_i -i[A_0,X_i] \ , \ \ \ \ \
{\cal F}_{ij}=-i[X_i,X_j] \  .
\ee
The  one-loop effective action is calculated using the background  field
method by the   standard procedure (see   \ci{dewit};  the
one-loop effective action for $D=10$ SYM   and its
   dimensional reductions   was  considered in \ci{fraa}).
 We decompose ${\cal
A}_{\mu}$ into the background and fluctuation parts,
\be
{\cal A}_{\mu}\rightarrow {\cal A}^{\prime}_{\mu}={\bar {\cal
A}}_{\mu}+{\cal A}_{\mu}=\LB
\begin{array}{cc}
 \bar{\cal A}_{\mu}^{(1)} & 0 \\
 0 & \bar{\cal A}_{\mu}^{(2)} \\
 \end{array}
 \RB + \LB
 \begin{array}{cc}
 0 & a_{\mu} \\
 a_{\mu}^{\dagger} & 0 \\
 \end{array}
 \RB  \,,
\ee
where  the indices (1) and (2) refer to the  two BPS objects
in the scattering problem of section 3.
 One needs  to   take  into account
only the off-diagonal fluctuations because the contribution
to the  effective action due to
self-interaction of BPS objects vanishes.
We  shall assume that
the
fermions $\theta$  and ghosts $B,C$ have vanishing background values  and
that  their
fluctuations are:
\be
\te  =  \LB
\begin{array}{cc}
 0 & \psi \\
 \psi^\T & 0 \\
 \end{array} \RB \,, \ \ \ \
B =  \LB
\begin{array}{cc}
 0 & b \\
 b^{\dagger} & 0 \\
 \end{array} \RB \,,\ \ \ \
C  =  \LB
\begin{array}{cc}
 0 & c \\
 c^{\dagger} & 0 \\
 \end{array} \RB \,,
\ee
where $\T$ is the matrix transposition.

The gauge-fixed action is
\be
L^{\prime}=\frac{1}{g_s}\tr \bigg[ -\frac{1}{4} {\cal
F}^{\prime}_{\mu\nu}{\cal F}^{\prime\mu\nu}-\frac{1}{2}(\bar{D}^{\mu}{\cal
A}_{\mu})^2\bigg]  +\hbox{fermionic term}+\hbox{ghost term} \,,
\ee
where
$
\bar{D}^{\mu}{\cal A}_{\mu}=\de^{\mu}{\cal A}_{\mu}-i[\bar{{\cal
A}}^{\mu},{\cal A}_{\mu}] .$
Using that
\be
\int dy^{\dagger}dy e^{-\tr( y^{\dagger}AyB)} =\bigg[\hbox{det}(A\otimes
B^\T)\bigg]^{-1}  ,
\ \ \ \ \
\int d\bar{\eta}d\eta e^{-\tr( \bar{\eta}A\eta B)} =\hbox{det}(A\otimes B^\T)
\ , \ee
for the bosonic and fermionic matrices respectively, ones  finds, after
a tedious but
straightforward
algebra (and the Wick rotation to the Euclidean time, 
$t\rightarrow i\tau ,A_0\rightarrow -iA_{\tau}$)  the  expressions
 for the
effective action given in section 3.1.

%%%%%%%%%%%%%%%%%%% EFFECTIVE ACTION OF GRAV.-FIVE BRANE %%%%%%%%%%%%%%%%%%%

\appendix{Super Yang-Mills  effective action  for  graviton--five brane
scattering}

The operator in \eq{wb1} is composed of 3 blocks in Lorentz space:
$2\times 2$ block $(b_n^2+P^2)+ E$, $4\times 4$ block $b_n^2+P^2$ and
$4\times 4$ block $b_n^2+P^2+2iG$. $2\times 2$ block can be diagonalised,
with the diagonal elements being $b_n^2+P^2\pm 2iv$.
We assume that the eigenvalues of $\gamma_9$ are $\pm 1$
with $\tr 1=8$ in the eigenspace.
Let us define
\be
c_n^2= b_n^2+2iv \,, \ \ \
d_n^2= b_n^2-2iv \,, \ \ \
e_n^2= b_n^2+iv \,,  \ \ \
f_n^2= b_n^2-iv \,.
\ee
The derivative of the effective action \rf{effact}
with respect to $b^2$ is given by
$$
\frac{d}{db^2}W=\frac{d}{db^2}W_{A=0}+ n_0 \sum^\infty_{n=0} \tr \bigg\{
\bigg[ \frac{1}{(P^2+b_n^2)}-\frac{1}{(P^2+b_n^2)_{A=0}}\bigg]
$$ $$
+ \
\bigg[ \frac{1}{(P^2+c_n^2)}-\frac{1}{(P^2+c_n^2)_{A=0}}\bigg]
+
\bigg[ \frac{1}{(P^2+d_n^2)}-\frac{1}{(P^2+d_n^2)_{A=0}}\bigg] $$ $$
 - \ 4 \
\bigg[
\frac{1}{(P^2+e_n^2)}-\frac{1}{(P^2+e_n^2)_{A=0}} \bigg]
-4
\bigg[
\frac{1}{(P^2+f_n^2)}-\frac{1}{(P^2+f_n^2)_{A=0}}\bigg]
$$ $$
+\
\bigg[\frac{1}{P^2+b_n^2}\sum_{k=2}^{\infty}\LB
\frac{1}{P^2+b_n^2}(-2iG)\RB^k\bigg]
-
\frac{1}{2}\bigg[ \frac{1}{P^2+e_n^2}\sum_{k=2}^{\infty}\LB
\frac{1}{P^2+e_n^2}(-\frac{i}{2}\gamma G)\RB^k\bigg]
$$ \be
-\ \frac{1}{2}\bigg[ \frac{1}{P^2+f_n^2}\sum_{k=2}^{\infty}\LB
\frac{1}{P^2+f_n^2}(-\frac{i}{2}\gamma G)\RB^k \bigg]\bigg\} \,.
\label{derW}
\ee
The traces in \eq{derW} can be calculated
as  expansions in  powers of $1/m^2$
(where  $m^2$ is
any of $b_n^2,c_n^2,d_n^2,e_n^2,f_n^2$)  using the operator Schwinger
method.\foot{A good review of the latter can be found in
\cite{novikov}. We thank K. Zarembo for pointing out that paper
to us.}
This  corresponds to   the long-distance ($b\to \infty$)
expansion  since  $b_n^2=b^2 + i v (2n+1)$, etc.

The large-mass  expansion of \rf{derW} is
\be
\frac{d}{db^2}W=\frac{d}{db^2}W_{A=0}+\frac{d}{db^2}I_0+
\frac{d}{db^2}I_1 +... \,,
\ee
where $dI_0/db^2$ comes from  terms proportional to $1/m^2$,
 $dI_1/db^2$ -- from  terms proportional to $1/m^4$, etc. Since we  are
interested in the leading order in the $1/b^2$ expansion,  only
the two terms $W_{A=0}$ and $I_0$ are relevant.
Because of supersymmetry (bose-fermi cancellation) all terms
in this expansion are UV convergent
(however, the IR divergences of the  1-loop YM effective action in external
fields  remain also in the supersymmetric case \ci{fraa}).

The basic  formulae are \ci{novikov}
\be
\tr\lsb \frac{1}{(P^2+m^2)}-\frac{1}{(P^2+m^2)_{A=0}}\rsb =
-\frac{1}{2^6\cdot 3\cdot \pi^2\cdot m^2}
\int d^4x{\ \rm tr} (G_{ab} G_{ab}) +O\LB \frac{1}{m^4} \RB\,,
\ee
\be
\tr\lsb\frac{1}{P^2+m^2}\LB
\frac{1}{P^2+m^2}(-\frac{i}{2}\gamma G)\RB^2\rsb =
\frac{1}{2^3\cdot \pi^2\cdot m^2}
\int d^4x{\ \rm tr} (G_{ab} G_{ab}) +O\LB \frac{1}{m^6} \RB \ ,
\ee
\be
\tr\lsb\frac{1}{P^2+m^2}\LB
\frac{1}{P^2+m^2}(-2iG)\RB^2\rsb =
\frac{1}{2^3\cdot \pi^2\cdot m^2}
\int d^4x{\ \rm tr} (G_{ab} G_{ab}) +O\LB \frac{1}{m^6} \RB\,.
\ee
In   general, in a   self-dual $G_{ab}$ background
$$
\tr\lsb\frac{1}{P^2+m^2}\LB
\frac{1}{P^2+m^2}(-2iG)\RB^k\rsb
=
\tr\lsb\frac{1}{P^2+m^2}\LB
\frac{1}{P^2+m^2}(-\frac{i}{2}\gamma G)\RB^k\rsb .
$$
This  relation  implies, in particular,  that the
$N=4, D=4$ SYM effective action vanishes in a self-dual
gauge field  background, i.e.  that in the present case
$W$ does not contain a static ($v=0$)  potential term.

We  find that
\be
W_{A=0}=n_0N_0\td V_4\sum^\infty_{n=0}{\ \rm Tr}_k\ \hbox{ln}\lsb
\frac{(k^2+b_n^2)^6(k^2+c_n^2)(k^2+d_n^2)}
{(k^2+e_n^2)^4(k^2+f_n^2)^4} \rsb \,,
\ee
where the factor $n_0N_0$ comes from the trace  $\tr (
{\bf 1}_{N_0\times N_0}\otimes {\bf 1}_{n_0\times n_0})$ and
${ \rm Tr}_k\ $ is the trace over discrete momenta on the torus
$\td T^4$.
Also,
\bea
\lefteqn{\frac{d}{db^2}I_0=n_0
 \int_{\td T^4}  d^4x{\ \rm tr} (G_{ab} G_{ab}) \ \sum^\infty_{n=0}\lsb
\frac{1}{2^4\cdot\pi^2}\LB\frac{2}{b_n^2}
-\frac{1}{e^2_n}-\frac{1}{f_n^2}\RB \right.}\non
&&- \left.
\frac{1}{2^6\cdot
3\cdot\pi^2}\LB\frac{6}{b^2_n}+\frac{1}{c_n^2}+\frac{1}{d_n^2}-\frac{4}{e_n^2}
-\frac{4}{f_n^2}\RB \rsb \, ,
\eea
$$
\frac{d}{db^2}I_1 =  in_0 \int_{\td T^4} d^4x{\ \rm tr} (
 G_{ab}G_{bc}
G_{ca}) $$
$$ \times
\sum^\infty_{n=0}\lsb
\frac{1}{2^5\cdot 3^2\cdot 5\cdot\pi^2}
\LB\frac{6}{b^4_n}+\frac{1}{c_n^4}+\frac{1}{d_n^4}-\frac{4}{e_n^4}
-\frac{4}{f_n^4}\RB +
\frac{1}{2^3\cdot
3\cdot\pi^2}\LB\frac{2}{b_n^4}-\frac{1}{e^4_n}-\frac{1}{f_n^4}\RB\rsb\ .
$$
Similarly, $\frac{d}{db^2}I_2$ is proportional to
 $\int d^4x{\ \rm tr}(  G^4 ) $, etc.

The  relations
\be
{\ \rm Tr}_k\bigg[\frac{1}{(k^2+m^2)^n}\bigg]
=\int \frac{d^4k}{(2\pi)^4}\frac{1}{(k^2+m^2)^n} + O\LB\frac{1}{m^{2n}}
\RB \,,
\ee
\be
 \ln \frac uv =\int\limits_0^{\infty}\frac{ds}{s}\left(e^{-vs}
- e^{-us} \right)
 \ , \ \ \ \ \ \
 \sum_{n=0}^\infty e^{-b_n^2s}=\frac{e^{-b^2s}}{2i\sin vs} \ ,
\ee
imply that
\be
W_{A=0}= - i \frac{n_0N_0\td V_4}
{16\pi^2  }\int_0^{\infty}\frac{ds}{s^3}e^{-b^2s}
\frac{1}{\sin vs}\LB 4\cos vs-\cos 2vs -3\RB + ... \ ,
\label{W0}
\ee
and
\be
I_0=\bigg(S_1 -  { 1 \ov 12} S_2\bigg)\frac{n_0}{16\pi^2}\int d^4x
 {\ \rm tr} (G_{ab}G_{ab}) \,,
\label{I0}
\ee
where we have defined
\be
S_1=-i\int_0^{\infty}\frac{ds}{s}e^{-b^2s}\frac{1}{\sin vs}(\cos vs
-1) \    =  \  O ({v\ov b^2})  \ ,
\label{S1}
\ee
and
\be
S_2=-i \int_0^{\infty}\frac{ds}{s}e^{-b^2s}\frac{1}{\sin vs}(4\cos vs
-\cos 2vs -3) \   = \   O ({v^3\ov b^6}) \,.
\label{S2}
\ee
Expanding  \rf{W0}, \rf{S1} and \rf{S2} in the powers of $1/b^2$ and
retaining only the leading term, we get the phase
shift~\rf{phase}.

The higher-order terms  coming from $I_1, I_2,...$, e.g., 
$$
I_1 = - { i n_0 \ov 24 \pi^2}
{d \ov d b^2 }  \bigg(S_1 -  { 1 \ov 60} S_2\bigg)  \int d^4x{\ \rm tr} (
 G_{ab}G_{bc}
G_{ca})\  =\  O( { v\ov b^4}) \  ,  $$
$$
I_2 =
\frac{n_0 }{2^9\cdot 3^2\cdot \pi^2}\bigg(  \frac{d^2}{d(b^2)^2}S_2\bigg)  \int
 d^4x {\ \rm tr} \bigg[(G_{ab} G_{ab})^2+\frac{1}{5}\{ G_{ab},
 G_{bc}\}^2 $$  $$  +\  \frac{1}{7}
[G_{ab}, G_{bc}]^2
+\frac{1}{70}[G_{ab},G_{cd}]^2 \bigg]
 \  =\   O( { v^3\ov b^{12}}) \  ,   $$
give only  subleading contributions.

%%%%%%%%%%%%%%%%%%%%%%%%%%%%%%%%%%%%%%%%%%%%%%%%%%%%%%%%%%%%%%%
%%%%%{\vspace*{\fill}\pagebreak}%%%%%%%%%%%%%%%%%%%%%%%%%%%%%%%%

{\vspace*{\fill}\pagebreak}


\begin{thebibliography}{99}
\addtolength{\itemsep}{-6pt}

\bibitem{bfss}
 T. Banks, W. Fischler, S.H. Shenker and L. Susskind,
{\it M Theory as a Matrix Model: a Conjecture}, hep-th/9610043.

\bibitem{wit1}
P.K. Townsend, \pl B350 (1995) 184, hep-th/9501068;
E. Witten, {Nucl.  Phys.} {B443} (1995) 85,
hep-th/9503124;
 C. Hull and P.K. Townsend, {Nucl. Phys.} {B438}
(1995) 109, hep-th/9410167;
 J.H. Schwarz, {\it Lectures on Superstring and
       M Theory Dualities}, hep-th/9607201;
       M.J. Duff, \ijmp  A12 (1997) 1215, hep-th/9608117; P.K.
       Townsend, {\it Four Lectures on M-theory},
hep-th/9612121.

\bibitem{pol}
J.~Polchinski, Phys.\ Rev.\ Lett. {75} (1995) 4724;
 {\it TASI Lectures on D-Branes}, hep-th/9611050.

 \bibitem{wit2}
E. Witten, {Nucl. Phys.} {B460} (1995) 335, hep-th/9510135.

\bibitem{tay}
W. Taylor,   {\it D-brane field theory on compact spaces},
hep-th/9611042.

\bibitem{sus}
L. Susskind,  {\it T Duality in M(atrix) Theory
and S Duality in field theory}, hep-th/9611164.


\bibitem{grt}
O.J.~Ganor, S.~Ramgoolam and W.~Taylor,
{\it Branes, Fluxes and
Duality in M(atrix) Theory}, hep-th/9611202.

\bi{dvv}
R. Dijkgraaf, E. Verlinde and  H. Verlinde,
{\it Matrix String Theory}, hep-th/9703030.



\bibitem{bss}
T.~Banks, N.~Seiberg and S.~Shenker,
{\it Branes from Matrices}, hep-th/9612157.
\bibitem{ab}
O. Aharony and M. Berkooz,
{\it Membrane Dynamics in M(atrix) Theory},
hep-th/9611215.

\bibitem{lifmat}
G. Lifschytz and S.D. Mathur,
{\it Supersymmetry and Membrane Interactions in M(atrix) Theory},
hep-th/9612087.


\bibitem{lif3}
G. Lifschytz,
{\it Four-Brane and Six-Brane Interactions in M(atrix) Theory},
hep-th/9612223.


\bibitem{corr}
D. Berenstein and R. Corrado, {\it M(atrix) theory in various dimensions},
hep-th/9702108.

\bibitem{pp}
J. Polchinski  and P. Pouliot,
{\it Membrane Scattering with M-Momentum Transfer},
 hep-th/9704029.


 \bi{nic}
 B. de Wit, J. Hoppe and H. Nicolai, \np B305 (1988) 545.


 \bibitem{rut} J.G. Russo and A.A. Tseytlin,
{\it  Waves, boosted branes
and BPS states in M theory},  hep-th/9611047.



\bibitem{doug} M.R. Douglas, {\it
 Branes within Branes},  hep-th/9512077.

\bi{ts4}
A.A.  Tseytlin,
{\it  Composite BPS configurations of p-branes in 10 and 11 dimensions},
 hep-th/9702163.


\bibitem{lif2} G. Lifschytz, {\it Probing bound states of D-branes},
hep-th/9610125.

\bibitem{dkps}
M.R. Douglas, D. Kabat, P. Pouliot and S.H. Shenker,
{\it  D-branes and Short Distances in String Theory},
hep-th/9608024.




\bi{guv}
 R. G\"uven, \pl B276 (1992) 49.


  \bi{ts2}
 A.A. Tseytlin, \np B475 (1996) 149,  hep-th/9604035.

 \bi{ts1}
 A.A. Tseytlin, \mpl A11 (1996) 689,  hep-th/9601177.

  \bibitem{wit3}
E. Witten, {Nucl. Phys.} {B460} (1995) 541, hep-th/9511030.

 \bibitem{lif4}
G. Lifschytz, {\it  A note on the transverse 5-brane in M(atrix) theory},
hep-th/9703201.

\bibitem{others}
E. Halyo, {\it A proposal for the wrapped transverse five-brane in M(atrix)
theory}, hep-th/9704086;
M. Berkooz and M. Rozali, {\it On the transverse fivebranes in M(atrix) theory on
$T^5$}, hep-th/9704089.

\bibitem{ballar}
V. Balasubramanian and F. Larsen,
{\it Relativistic brane scattering},
hep-th/9703039.


 \bi{calkhu} C.G. Callan and R.R. Khuri, \pl
 B261 (1991) 363.

 \bi{duff}
M.J. Duff, R. Khuri  and J.X. Lu, Phys. Rept.
 259 (1995) 213, hep-th/9412184.


  \bi{ts3}
 A.A. Tseytlin,
 {\it No-force condition and BPS combinations of p-branes in 11
 and 10 dimensions}, \np B487 (1997) 141,    hep-th/9609212.


 \bi{dps}
 M. Douglas,  J. Polchinski  and A.  Strominger,
 {\it Probing Five-Dimensional Black Holes with D-Branes},
 hep-th/9703031.


\bi{lei}
R.G. Leigh, \mpl A4 (1989) 2767.

\bi{FT}
 E.S. Fradkin and A.A.  Tseytlin, \pl B163 (1985) 123.

\bibitem{bachas}
C.  Bachas, Phys. Lett. {B374} (1996) 37,   hep-th/9511043.


\bi{ts5}
A.A. Tseytlin,
Nucl. Phys. B469 (1996) 51, hep-th/9602064.


%%%%%%%%%%%%%%%%%%%%%%%%%%%%%%%%%%%%%%%%%%%%%%%%%%%%%%%%%%%%%%%


\bi{HS} G.T.~Horowitz and A.~Strominger, Nucl. Phys. { B360}
(1991) 197.

\bibitem{lif1}
G. Lifschytz,
{\it Comparing D-branes to black holes},
hep-th/9604156.

 \bi{vbaal}
 P. van Baal, \cmp 94 (1984) 397.

 \bi{hashi}
 A. Hashimoto and W. Taylor,
 {\it Fluctuation Spectra of Tilted and Intersecting D-branes from the Born-Infeld
       Action},
       hep-th/9703217.

 \bi{polchi} J. Polchinski, \pl B209 (1988) 252;
J. Gauntlett, J. Harvey, M. Robinson and D. Waldram, \np B411 (1994) 461, 
hep-th/9305066;
C. Callan, J. Maldacena and A. Peet, \np B475 (1996) 645, hep-th/9510154.


 \bi{shira} K. Shiraishi, {\it Extreme dilatonic black holes on a torus},
 gr-qc/9511005;
  R. Khuri,
 Nucl. Phys. B403 (1993) 335;
 R.R. Khuri and R.C. Myers, {\it Low-energy scattering of black holes and
 p-branes in string theory},
 hep-th/9508045.


\bi{dewit}
B.S. De Witt, {\it Dynamical Theory of Groups and Fields},
Gordon and Breach, N.Y., 1965; G. 't Hooft,
Nucl. Phys. B62 (1973) 444.

\bi{fraa} E.S. Fradkin and A.A. Tseytlin, \np B227 (1983) 252.


\bi{novikov}V.A. Novikov, M.A. Shifman, A.I. Vainshtein
 and  V.I. Zakharov, {\it Calculations in external fields
 in Quantum Chromodynamics: technical review (abstract operator
 method, Fock-Schwinger gauge)},
 Fortsch. Phys. 32  (1985) 585.

\end{thebibliography}
\end{document}